\documentclass{SCGE}
\newcommand{\mpch}{\,h^{-1}{\rm {Mpc}}}
\newcommand{\lcdm}{\Lambda{\rm CDM}}

\let\citedash\relax
\makeatletter \providecommand{\citedash}{\hbox{-}\penalty\@m}
\makeatother

\begin{document}
%%%%%%ÐÂ°æÊ½Òª¼ÓÉÏÕâ×é
\begin{picture}(0,0){\rm
\put(0,-20){\makebox[160truemm][l]{\bf {\sanhao\raisebox{2pt}{.}}
Article  {\sanhao\raisebox{1.5pt}{.}}}}}
\put(0,-34){\jiuwuhao {\textcolor[rgb]{0.0,0.0,0.0}{\sf Special Topic: Fluid Mechanics
}}}%%(11ÔÂ×¢ÊÍ£ºµ÷\textcolor[rgb]{x,x,x}ÖÐµÄÊý×ÖxÔ½´óÔ½»Ò)
\end{picture}

\def\bm{\boldsymbol}

\def\dl{\displaystyle}
\def\du{\end{document}}
\def\d{{\rm d}}
\def\e{{\rm e}}
\def\i{{\rm i}}

% The author doesn't need fill in it.
\Year{2016} %
\Month{January} %
\Vol{59} %  ¾íºÅ
\No{1} %  ÆÚºÅ
\BeginPage{1} % ÆðÒ³Âë
\AuthorMark{{\rm A. Author}, et al.}  %(11ÔÂ×¢ÊÍ£ºÒ³Ã¼ÉÏµÄ×÷Õß)
\AuthorMarkCite{A. Author, B. Author, C. Author, and D. Author} %(11ÔÂ×¢ÊÍ£ºcitationÖÐµÄ×÷Õß)
\DOI{10.1007/s11433-015-5649-8} % The author doesn't need fill in it.
\ArtNo{000000}

% \title[short text for running head]{full title}{comments for title}
\title[CosmicGrowth Simulations)]{ CosmicGrowth Simulations---Cosmological simulations for
  structure growth studies }%±êÌâ {ÓÐºÚÌåµÄÊ±ºòÐèÒª½«±êÌâ¸´ÖÆÔÚÖÐÀ¨ºÅÀïÃæ£¬Ê¹µÃÒýÓÃÌõÏÔÊ¾°×Ìå¡£Ã»ÓÐºÚÌåµÄÊ±ºòÖÐÀ¨ºÅ¿ÉÒÔÉ¾µô}

\author[1,2*]{Y.P. Jing (景益鹏)}{}
\footnote{*Corresponding author (email: ypjing@sjtu.edu.cn)}%ÊÖ¶¯E-mailµØÖ·

\address[{\rm1}]{Department of Astronomy,  Shanghai Jiao Tong University, 800 Dongchuan
  Road, Shanghai, 200240,China}
\address[{\rm2}]{IFSA Collaborative Innovation Center, Shanghai Jiao Tong University, Shanghai 200240, China}
\address[{\rm3}]{Tsung-Dao Lee Institute, Shanghai Jiao Tong University, 800 Dongchuan
  Road, Shanghai, 200240,China}

\maketitle \vspace{-3.5mm}{\footnotesize\begin{center} Received January 1, 2016; accepted January 1, 2016; published online January 1, 2016%ÊÕ¸åÈÕÆÚ
\end{center}}\vspace*{-5mm}

%     Abstract is required.
\begin{center}
\rule{16.5cm}{0.4pt}
\parbox{16.5cm}
{\begin{abstract} 

I present a large set of high resolution simulations, called
CosmicGrowth Simulations, which were generated with either 8.6 billion
or 29 billion particles. As the nominal cosmological model that can
match nearly all observations on cosmological scales, I have adopted a
flat Cold Dark Matter (CDM) model with a cosmological constant
$\Lambda$ ($\lcdm$). The model parameters have been taken either from
the latest result of the WMAP satellite (WMAP $\lcdm$) or from the first
year's result of the Planck satellite (Planck $\lcdm$).  Six
simulations are produced in the $\lcdm$ models with two in the Planck
model and the others in the WMAP model. In order for studying the
nonlinear evolution of the clustering, four simulations were also
produced with $8.6$ billion particles for the scale-free models of an
initial power spectrum $P(k)\propto k^n$ with $n=0$,$-1$,$-1.5$ or
$-2.0$. Furthermore, two radical CDM models (XCDM) are simulated with
8.6 billion particles each. Since the XCDM have some of the model
parameters distinct from those of the $\lcdm$ models, they must be
unable to match the observations, but are very useful for studying how
the clustering properties depend on the model parameters. The
Friends-of-Friends (FoF) halos were identified for each snapshot and
subhalos were produced by the Hierarchical Branch Tracing (HBT)
algorithm. These simulations form a powerful database to study the
growth and evolution of the cosmic structures both in theories and in
observations.
  \end{abstract}}
\end{center}\vspace*{-0.6cm}

\begin{center}
\parbox{16.5cm}
{\bf\jiuhao galaxy formation, structure formation in the Universe, cosmology}%¹Ø¼ü´Ê
\end{center}

\begin{center}
{}%·ÖÀàºÅ(3~5 codes)  http://phys.scichina.com:8083/sciGe/UserFiles/File/pacs.pdf
\CITA    %%(11ÔÂ×¢ÊÍ£ºCitationÄÚÈÝ×Ô¶¯Éú³É)
%\Cit{~~~???, et al. ???. Sci China-Phys Mech Astron, 2014, 57: 1--6, doi:}%%(11ÔÂ×¢ÊÍ£ºCitationÄÚÈÝÐèÊÖ¶¯ÌîÐ´)
\end{center}

\textwidth=178truemm \textheight=236truemm%%%%%%ÐÂ°æÊ½Òª¼ÓÉÏ

%%%%%%%%%%%%%%%%%%%%%%%%%%%%%%%%%%%%%%%%%%%%%%%%%%%%%%%%%%%%
\wuhao\vspace*{1.5mm}

\begin{multicols}{2}

%%%%%%%%%%%%%%%%%%%%%%%%%%%%%%%%%%%%%%%%%%%%%%%%%%%%%%%%%%%%
%% Text of article.
%%%%%%%%%%%%%%%%%%%%%%%%%%%%%%%%%%%%%%%%%%%%%%%%%%%%%%%%%%%%
%    Section headings
\renewcommand{\baselinestretch}{1.08} \baselineskip 12.2pt\parindent=10.8pt

\renewcommand{\thefootnote}

\section{Introduction}\label{sec:1}%"sec:intro"ÊÇ¸ø¸Ã½ÚµÄÃüÃû£¨¿É»»ÆäËû£©£¬ÒýÓÃÊ±ÓÃ\ref{sec:intro}

In the past forty years N-body simulations have played a key role in
advancing our knowledge about structure formation in the
Universe(\cite{ref1,ref2} for recent reviews). Perturbation theories
have been tested, refined and calibrated with N-body simulations
\cite{ref3,ref4}. The clustering and abundance of dark matter halos
has been found to have discrepancies with the Press-Schechter theory
\cite{ref5,ref6}, which promoted massive studies on the
subject\cite{ref7,ref8}. Now the halo bias, including its dependence
on the halo assembly history\cite{ref9}, has been understood
reasonably well\cite{ref10}. The density profile of halos has been
found to be well described by the Navarro, Frenk and White
form\cite{ref11}. Furthermore, the internal structures are found to
rely on their mass growth\cite{ref12,ref13}, both of which can be
accurately described by simple scaling forms\cite{ref14,ref15}. The
halos have been found to be triaxial, and the shape distributions are
accurately given in simple scaling forms\cite{ref16}. The subhalo
abundance in each halo is approximately universal once the subhalo
mass is scaled by the host halo mass\cite{ref17,ref18}. All these
results have formed important ingredients for understanding how the cosmic
structures have developed. They are also the bases from which galaxy
formation theories are constructed and confronted with various
observations. N-body simulations are widely used to plant galaxies,
through so-called semi-analytical
modeling\cite{ref19,ref20,ref21,ref22}, to form model galaxy catalogs
that are used to understand various aspects of galaxy formation, and
are directly compared with observations of galaxies. N-body
simulations are also used to empirically derive the occupation
distributions, e.g. the luminosity function and the mass function of
certain type galaxies in dark matter halos, from large galaxy surveys
through adopting methods, such as a Halo Occupation Distribution
(HOD)\cite{ref23,ref24}, Conditional Luminosity Function\cite{ref25},
or Abundance Matching\cite{ref26,ref27,ref28}.

With the development of large galaxy surveys in the near future, the
properties of the dark energy, test of General Relativity, and the
formation and evolution of galaxies will become the focuses of the
future research. These studies have raised new demands for N-body
simulations. For example, for studying the dark energy and testing
gravity models, weak lensing and redshift distortion are two important
observables. However, to extract the physical information from the
observations, one has to take into account of all non-linear effects,
such as non-linear evolution of cosmic structures, spatial and
velocity biases of galaxies, intrinsic alignments of galaxies, real
space to redshift space mapping. This means that N-body simulations
have to well resolve galaxies, while they have to have a volume
sufficiently large to cover large scale structures in the Universe.

In this paper, I present a new set of N-body cosmological simulations
called CosmicGrowth simulations. Since the simulations have been
designed mainly for studying the acceleration of the cosmic expansion
and for studying the clustering of dark matter and galaxies, I have
generated simulations not only for ${\Lambda}$CDM (cosmological
constant $\Lambda$ plus Cold Dark Matter) models, but also for the
scale-free (SF) models and other CDM like models (XCDM) that have the
model parameters that are distinctly different from the
observed ones. The main reason for producing SF and XCDM simulations is
that they help us understand the nonlinear processes, formation of
galaxies, and the effects of the cosmological parameters. This also
manifests the uniqueness of our simulations that are distinct from
those of other groups who usually only do $\Lambda$CDM
simulations. As was shown by \cite{ref6,ref15}, the SF simulations are
very powerful for understanding the gravitational processes in forming
cosmic structures, including large scale structures and internal
structures of dark matter halos.

The paper is arranged as the following. In sect \ref{sec:2}, I will
describe the simulations, including how the catalogs of the dark
matter halos and subhalos are constructed. Then in sect \ref{sec:3}, I present the
mass functions of dark matter halos in a typical
$\Lambda$CDM model, to demonstrate the convergence of the halo catalog
in different simulations. Our main conclusions will be summarized in sect
\ref{sec:4}

\section{Simulations}\label{sec:2}

\subsection{Models}
I choose the cosmological parameters compatible with the WMAP
observations. I assume the universe is flat, with the current cosmic
density contributed by the vacuum energy (or equivalently the
cosmological constant) $\Omega_{\Lambda}$, by cold dark matter
$\Omega_{\rm c}$ and by baryonic matter $\Omega_{b}$, where
$\Omega_{x}$ is the density parameter, i.e. the density of the
corresponding component $x$ in units of the critical density of the
Universe at the current epoch. The primordial fluctuation is a
Gaussian one with a power-law spectrum $\propto k^{n_s}$, and its
amplitude is set by the {\it rms} linear density fluctuation
$\sigma_8$ in a sphere of radius $8\mpch$ at the current epoch, where
$h$ is the Hubble constant in terms of $100 {\rm \ km s^{-1}
  Mpc^{-1}}$ . In addition to the cosmic background radiation, there
are three species of massless neutrinos in the Universe. I choose the
cosmological model parameters as listed in Table 1, which are well
consistent with the Seven-Year data and Nine-Year data of WMAP
\cite{ref29,ref30}. I call this model as the WMAP cosmology.

After I had started the simulations, the Planck team released
their observational results that yield new constraints on the
cosmological parameters. While their results are consistent with
those of the WMAP at 2$\sigma$ level, I generate simulations for the
cosmological model set by their first year's data\cite{ref31}, in order to study how
the change of the parameters will impact on the structure formation . The cosmological
parameters are listed in Table 1 with a label {\rm Planck}.

While there is consensus that the $\Lambda$CDM models with the
parameters taken above are the best-fitting models to the observations, it is
known that there exist resolution problems in any
simulation that may cause difficulties in interpretation of the
simulation results especially on small or on the largest scales. The SF
simulations, because of its scaling properties by construction, may
help one understand the simulation limitations\cite{ref32}. Also because of its
scaling properties, it helps understanding the universal
properties of clustering, such as the non-linear clustering, and
the assembly, structures and clustering
of dark matter halos. It also helps improving theoretical
understanding of the halo formation, such as the seminal (Extended)
Press-Schechter theories. I believe that the scale-free simulations
will continue to play an important role in understanding the weak
lensing and redshift space distortion in the era of studying the cosmic
expansion. Therefore I have constructed SF simulations for $n=0$, $-1$,
$-1.5$, and $-2$ respectively (Table 2). Why I have chosen to run more
simulations around $n=-1.5$, because they have the slope similar to
that of $\Lambda$CDM at the critical scale where the clustering become
non-linear after redshift $z=3$. The initial condition is set in the
same way as \cite{ref6}, i.e. the initial power spectrum at the
particle Nyquist wavelength is $A/N_p$, where $N_p$ is the particle
number in the simulation and $A$ is listed Table 2. As demonstrated in
\cite{ref6}, in the case of $n=-2$, the fluctuation is quite nonlinear
even at the start of the simulation. I actually generate the initial
condition for $n=-2$ at $a=a_i/10$ with the fluctuation amplitude set
to $A/100$, and evolve the initial condition by the PM version to $a_i$.

I also choose two CDM-like cosmological models that have model
parameters strongly different that of the WMAP $\Lambda$CDM model. I
call them as the $X$CDM models (Table 3). The $X$CDM1 has the same
linear power spectrum as the WMAP $\lcdm$, but the universe is made of
CDM only with $\Omega_c=1$. The $X$CDM2 has the same density
parameters as the $X$CDM1 model, but its primordial power spectrum has
$n_s=1$ and the transfer function is adopted from Barden et
al. (BBKS)\cite{ref33} with $\Omega_c h=0.5$. While we know these
models do not fit the real Universe well, because of the change of the
model parameters, they may help us understand how the properties of
the structure growth will change with the parameters.

\begin{table*}
\caption{The model parameters of  $\lcdm$ simulations}\label{tab:table1}
\vspace{-1mm}\footnotesize
\begin{center} \doublerulesep 0.2pt \tabcolsep 10pt
\begin{tabular}{lcccccc}
\hline
Model & $\Omega_b$ & $\Omega_c$ & $\Omega_\Lambda$ & $h$ & $n_s$ &$\sigma_8$\\

 \hline
Planck $\lcdm$ & $0.0487$&$0.2663$&$0.685$&$0.673$&$0.9603$&$0.829$  \\
WMAP $\lcdm$ & $0.0445$&$0.2235$&$0.732$&$0.71$&$0.968$&$0.83$  \\
 \hline
\end{tabular}
\end{center}
\end{table*}

\begin{table*}
\caption{The model parameters of the SF simulations}\label{tab:table2}
\vspace{-1mm}\footnotesize
\begin{center} \doublerulesep 0.2pt \tabcolsep 10pt

\begin{tabular}{lcccc}
\hline
Model & $\Omega_c$ & $\Omega_\Lambda$ & $n$&$A$\\

 \hline
SFn0 &$1$&$0$&$0$&$0.7314$  \\
SFn-1 &$1$&$0$&$-1$&$0.5629$  \\
SFn-1p5 &$1$&$0$&$-1.5$&$0.4891$  \\
SFn-2 &$1$&$0$&$-2.0$&$0.4241$  \\

 \hline
\end{tabular}
\end{center}
\end{table*}

\begin{table*}
\caption{The model parameters of the XCDM simulations}\label{tab:table3}
\vspace{-1mm}\footnotesize
\begin{center} \doublerulesep 0.2pt \tabcolsep 10pt
\begin{tabular}{lcccccc}
\hline
Model & $\Omega_b$ & $\Omega_c$ & $\Omega_\Lambda$ & $n_s$ &$\sigma_8$ &linear $P(k)$\\

 \hline

XCDM1 & $0.0$&$1.0$&$0$&$0.968$&$0.83$ & WMAP $\lcdm$ \\
XCDM2 & $0.0$&$1.0$&$0$&$1.0$&$0.83$ &BBKS \\
 \hline
\end{tabular}
\end{center}
\end{table*}

\subsection{Simulation Parameters} 
For $\Lambda$CDM and $X$CDM models, I choose an initial redshift
$z_i$ at which the simulation starts.  The simulations are evolved
with my adaptive parallel ${\rm P^3M}$ N-body code. The force between
the particles is softened at small scale. The S2 form of Efstathiou et
al \cite{ref34} is taken, with the softening parameter $\eta$ being
the scale beyond which the force between two particles is exact. This
parameter is a key one that determines the simulation resolution at
small scale. It also sets a constraint on the time step. Here I choose
a constant time step $da$ in the universe scale factor $a$. Guided by
our previous simulations, I set $\eta$, $da$ (and hence $N_{step}$)
as listed in Table 4 and Table 6 for $\lcdm$ and $X$CDM simulations
respectively. The scale factor $a$ is normalized
to $a_i=1$ at $z_i$, and the box size $L$ and the softening length
$\eta$ are in unit of $\mpch$. In order to identify subhalos and form
merger histories, I have output a large number of snapshots $N_{\rm
  snap}$ with an equal logarithmic interval in the scale factor $a$
from $z_{\rm out,1}$ to $z=0$.

Following \cite{ref32}, I take the time variable $p=a^{\frac{3}{n+3}}$
for the SF model with index $n$. The $\eta$ parameter is in units of
the box size $L$. I evolve the the simulation to an epoch when the rms
density fluctuation of a sphere of radius $R=0.05 L$ becomes
non-linear according to the linear perturbation theory. The simulation
parameters are listed in Table 5. I start the simulation at $p_i=1$,
and output the snapshots from $p_{out,1}$ in an equal logarithmic
interval of $p$. The meaning of the other simulation parameters are
the same as those for the $\lcdm$ models.

For XCDM models, I take the same simulation parameters as I did for
WMAP\_2048\_1200 or Planck\_2048\_1200 (see Table 6).

\begin{table*}
\caption{The $\lcdm$ simulations and the simulation parameters}\label{tab:table1}
\vspace{-1mm}\footnotesize
\begin{center} \doublerulesep 0.2pt \tabcolsep 10pt
\begin{tabular}{lcccccccccc}
\hline
Name & model & $N_p$ & $L$ & $da$&$\eta$&$z_i$ & $N_{step}$ &$z_{out,1}$&$N_{snap}$&realizations\\

 \hline
Planck\_2048\_400&Planck $\lcdm$ & $2048^3$&$400$&$0.0288$&$0.007$&$144$&$5000$&$16.87$&$100$&1  \\
Planck\_2048\_1200&Planck $\lcdm$ & $2048^3$&$1200$&$0.06$&$0.03$&$72$&$1200$&$7.30$&$24$&1  \\
WMAP\_2048\_400&Planck $\lcdm$ & $2048^3$&$400$&$0.0288$&$0.007$&$144$&$5000$&$16.87$&$100$&1  \\
WMAP\_2048\_1200&Planck $\lcdm$ & $2048^3$&$1200$&$0.06$&$0.03$&$72$&$1200$&$7.30$&$24$&1  \\
WMAP\_3072\_600&Planck $\lcdm$ & $3072^3$&$600$&$0.0288$&$0.01$&$144$&$5000$&$16.87$&$100$&3  \\
WMAP\_3072\_1200&Planck $\lcdm$ & $3072^3$&$1200$&$0.0288$&$0.02$&$144$&$5000$&$16.87$&$100$&1  \\
 \hline
\end{tabular}
\end{center}
\end{table*}

\begin{table*}
\caption{The SF simulations and the simulation parameters}\label{tab:table1}
\vspace{-1mm}\footnotesize
\begin{center} \doublerulesep 0.2pt \tabcolsep 10pt
\begin{tabular}{lccccccccc}
\hline
Name & model & $N_p$ & $dp$&$\eta$&$z_i$ & $N_{step}$ &$p_{out,1}$&$N_{snap}$&realizations\\

 \hline
SFn0\_2048& SFn0 & $2048^3$&$0.0202$&$0.125\times 10^{-4}$&$1029.15$&$5000$&$2.37$&$50$&2  \\
SFn-1\_2048& SFn-1 & $2048^3$&$0.0302$&$0.125\times 10^{-4}$&$151.00$&$5000$&$3.54$&$50$&2  \\
SFn-1p5\_2048& SFn-1p5 & $2048^3$&$0.0401$&$0.125\times 10^{-4}$&$52.51$&$5000$&$4.69$&$50$&2  \\
SFn-2\_2048& SFn-2 & $2048^3$&$0.0602$&$0.125\times 10^{-4}$&$16.38$&$5000$&$7.02$&$50$&1  \\
 \hline
\end{tabular}
\end{center}
\end{table*}

\begin{table*}
\caption{The XCDM simulations and the simulation parameters}\label{tab:table6}
\vspace{-1mm}\footnotesize
\begin{center} \doublerulesep 0.2pt \tabcolsep 10pt
\begin{tabular}{lcccccccccc}
\hline
Name & model & $N_p$ & $L$ & $da$&$\eta$&$z_i$ & $N_{step}$ &$z_{out,1}$&$N_{snap}$&realizations\\

 \hline
XCDM1\_2048\_1200&XCDM1 & $2048^3$&$1200$&$0.06$&$0.03$&$72$&$1200$&$7.30$&$24$&1  \\
XCDM2\_2048\_1200&XCDM2 & $2048^3$&$1200$&$0.06$&$0.03$&$72$&$1200$&$7.30$&$24$&1  \\
 \hline
\end{tabular}
\end{center}
\end{table*}

\subsection{FoF Groups, halos,  and subhalos}

Groups are identified with the Friends-of-Friends (FoF) algorithm with
the linking length taken to be 0.2 times of the mean particle
separation, and a FoF group catalog is constructed for each snapshot.
It is known that some of the member particles in a FoF group are not
bound, and the fraction of unbound particles increases with the
decrease of the group mass. Following the practice of \cite{ref35}, I
have removed all unbound particles from the FoF groups, and form a
bound halo catalog for each snapshot. As one will see in the next
section, the mass function of the halos converges well at the halo
mass equivalent to $\sim 10$ times of the particle mass, while the
effect of the unbound particles can be seen clearly for original FoF
groups even at halo mass equivalent to 100 times of particle
mass\cite{ref35}. Therefore, I recommend to use the bound halo
catalogs for future statistical studies.

Halos falling into a more massive halo will become subhalos. Subhalos
are interesting structures, because they are the hosts of satellite
galaxies in groups or clusters of galaxies. They also form potential
wells to confine gas and to support the motion of the stars inside. 
The Hierarchical Branch Tracing  (HBT) code\cite{ref36} has been used to
construct the merger tree and the subhalo catalogs.

\section{Mass Functions of Dark Matter Halos}\label{sec:3}
In Figure 1, I present the mass functions of the bound FoF halos in
the WMAP cosmology at z=0. The red dots are for the simulation
WMAP\_3072\_1200, and the blue and red lines are respectively for two
realizations of the simulation WMAP\_3072\_600. I have plotted the
mass functions above the halo mass corresponding to 10 particles. Since
WMAP\_3072\_600 has resolutions 8 times better in mass and 2 times
better in force than WMAP\_3072\_1200, it is amazing to see
that mass functions agree nearly perfectly at the mass $5\times 10^{10}
{\rm M_\odot}h^{-1}$ which corresponds to the mass of 13 particles in
WMAP\_3072\_1200. Furthermore, one can see the mass functions of the two
realizations of WMAP\_3072\_600 perfectly agree on the whole mass range
except for the very massive end where fluctuations are expected for
different realizations. The plot indicates that our procedure to remove
the unbound particles in the FoF groups are very successful, and the
bound group catalogs may be used for statistical analyses for the halo
members above $\sim 13$.

\begin{figure}[H]
\includegraphics[scale=0.40]{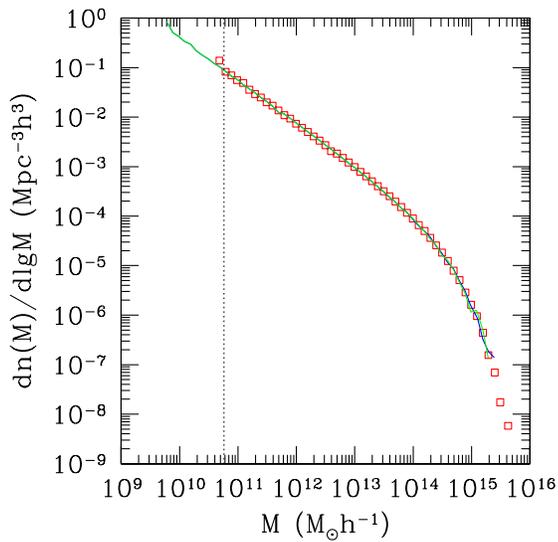}
\caption{The mass functions of the halos in WMAP\_3072\_1200 (red
  dots) and in two realizations (blue and green lines respectively) of
  WMAP\_3072\_600.} 
\label{fig:1}
\end{figure}

\section{Discussion and conclusions}\label{sec:4}
I present a large set of high resolution simulations which were
generated with either 8.6 billion or 29 billion particles. As for the
nominal cosmological model that can match nearly all observations on
cosmological scales, I have adopted a flat Cold Dark Matter (CDM)
model with a cosmological constant $\Lambda$ ($\lcdm$). The model
parameters have been taken either from the latest result of the WMAP
satellite (WMAP $\lcdm$) or from the first year's result of the Planck
satellite (Planck $\lcdm$).  Six simulations are produced in the
$\lcdm$ models with two in the Planck model and the others in the WMAP
model. In order for studying the nonlinear evolution of the
clustering, four simulations were also produced with $8.6$ billion
particles for the scale-free models of an initial power spectrum
$P(k)\propto k^n$ with $n=0$,$-1$,$-1.5$ or $-2.0$. Furthermore, two
radical CDM models (XCDM) are simulated with 8.6 billion particles
each. Since XCDM models have some of the model parameters distinct
from those of the $\lcdm$ models, they must be unable to match the
observations, but are very useful for studying how the clustering
properties depend on the model parameters. The Friends-of-Friends
(FoF) halos were identified for each snapshot and subhalos were
produced by the Hierarchical Branch Tracing (HBT) algorithm. I have
demonstrated that the halos are well resolved and identified at a mass
above that of 13 particles. These simulations form a powerful database
to study the growth and evolution of the cosmic structures both in
theories and in observations.

\vspace*{2mm} \Acknowledgements{\bahao } The work is supported by the
NSFC (11320101002, 11533006, \& 11621303) and 973 Program
No. 2015CB857003. I am very grateful to Jiaxin Han for identifying
subhalos for the $\Lambda$CDM simulations.

\end{multicols}
\end{document}